\documentclass{article}

\usepackage{PRIMEarxiv}

\usepackage[utf8]{inputenc} 
\usepackage[T1]{fontenc}    
\usepackage{hyperref}       
\usepackage{url}            
\usepackage{booktabs}       
\usepackage{amsfonts}       
\usepackage{nicefrac}       
\usepackage{microtype}      
\usepackage{lipsum}
\usepackage{fancyhdr}       
\usepackage{graphicx}       
\graphicspath{{media/}}     

\usepackage{mathrsfs,amsmath} 
\usepackage{mathtools}  
\usepackage{amsfonts}  
\usepackage{adjustbox}
\usepackage{diagbox}

\usepackage{algorithm}
\usepackage{algpseudocode} 
\usepackage{appendix} 

\usepackage{xcolor}
\usepackage{soul}

\title{Adaptive Surrogate Modeling for High-Dimensional Spatio-Temporal Output
\thanks{\textit{\underline{Citation}}: 
\textbf{Kapusuzoglu, B., Matsumoto, S., Miyagi, Y., Watanabe, D., \& Mahadevan, S. Adaptive surrogate modeling for high-dimensional spatio-temporal output. \textit{Structural and Multidisciplinary Optimization} 65, 290 (2022). DOI: \href{https://doi.org/10.1007/s00158-022-03402-x}{10.1007/s00158-022-03402-x}}
} 
}

\author{
  \textbf{Berkcan Kapusuzoglu}\thanks{Corresponding author: berkcan.kapusuzoglu@vanderbilt.edu}$^{\ ,1}$, \textbf{Shunsaku Matsumoto}$^2$, \textbf{Yoshitomo Miyagi}$^3$, \\ \textbf{Daigo Watanabe}$^2$, \textbf{Sankaran Mahadevan}$^1$ \\
  \vspace{0.2cm} \\
  $^1$Department of Civil and Environmental Engineering, Vanderbilt University, Nashville, TN 37235, USA \\
  $^2$Strength Research Department, Research and Innovation Center, Mitsubishi Heavy Industries, Ltd., \\ Nagasaki, 851-0392, Japan \\
  $^3$Strength Research Department, Research and Innovation Center, Mitsubishi Heavy Industries, Ltd., \\ Takasago, 676-8686, Japan
}

\begin{document}
\maketitle

\begin{abstract}
This paper develops an adaptive surrogate modeling method for problems with very high-dimensional spatio-temporal outputs. The analysis of spatio-temporal multi-physics systems is computationally expensive and consists of a large number of inputs and outputs. Surrogate models are often constructed to replace the physics-based model to achieve computational efficiency in analyses such as uncertainty quantification and optimization that require many function calls. In order to address the challenge introduced by the high dimensionality of spatio-temporal output, a dimension reduction method is first employed to map the high-dimensional output to a low-dimensional latent space. This is followed by the construction of the surrogate model in the low-dimensional space. The prediction error in the original space, which includes both the reconstruction error and surrogate model error, is evaluated using different error metrics. Based on the prediction accuracy of the surrogate model, new training points are identified for adaptive improvement of the surrogate model. We present a novel adaptive sampling technique that combines exploration and exploitation to improve the surrogate model accuracy with the fewest possible runs of the expensive physics-based model. Thermo-mechanical analysis of a gas turbine engine blade is used to analyze the effectiveness of the proposed method.
\end{abstract}





\section{Introduction}\label{sec:Intro}

Engineering analyses such as design optimization, reliability assessment, and system health diagnosis and prognosis often require techniques such as uncertainty quantification, model calibration, and optimization, which require multiple runs of the physics model. Computational models are often used to analyze the response of an engineering system for a variety of input realizations, since conducting experiments to directly measure the true response for many input realizations is often not affordable. For large mechanical systems, expensive finite element-based or computational fluid dynamics-based physics models are commonly employed. When it is not affordable to run the expensive physics-based model many times, an inexpensive surrogate model becomes necessary to carry out repetitive analyses such as design optimization, uncertainty quantification (UQ), probabilistic diagnosis and prognosis, and risk analysis to support the decision making \cite{berkcan2020Optimization,berkcan_2022multi}. The construction of an accurate surrogate model requires adequate amount of training data that can be generated by evaluating the physics-based model at multiple settings in the input space.

The challenge regarding computational effort becomes even greater in the case of nonlinear behavior under extreme environments, as in gas turbine engine components, where the structural response is governed by a coupled multidisciplinary system of equations (fluid mechanics, heat transfer, and structural mechanics) with high-dimensional system outputs that vary over space and time. The quantity of interest (QoI) is often a multivariate output that is a field quantity (exhibits spatial dependence) and/or a stochastic process (exhibits temporal dependence). Surrogate models are often needed in such cases as mentioned above, but the surrogate model prediction quality is highly dependent on the size and distribution of the training data, which is obtained by running the original physics-based model at different input settings. For nonlinear systems with high-dimensional output, such repeated evaluations pose significant challenge w.r.t. computational resources and time, thus motivating the minimization of the number of training runs of the expensive physics-based model.

Several types of surrogate models are used in the literature, e.g., polynomial chaos expansion (PCE) \cite{xiu2002wiener}, Gaussian process (GP) regression \cite{Rasmussen2004}, support vector regression (SVR) \cite{cortes1995support}, deep neural networks (DNNs), etc. The first two approaches PCE and GP are computationally demanding in problems with high-dimensional outputs. Co-Kriging~\cite{myers1982matrix} has been used for multivariate outputs, but it can be computationally demanding for high-dimensional field outputs~\cite{gogu2013efficient}. As for SVR and DNN and other machine learning models, the accuracy is dependent on the quality and quantity of the training data, which is often limited if the physics model runs are expensive \cite{ress_infoFusion_bk}.

Several studies on surrogate modeling techniques have addressed high-dimensional output by mapping the output to the space of principal directions, where the top few components capture most of the variance in the output, using methods such as principal components analysis (PCA) \cite{WOLD198737,hombal2013surrogate}, or singular value decomposition (SVD))~\cite{nath2017sensor, guo2021surrogate,hu2017surrogate}. PCA and SVD are challenging for very high-dimensional problems since they require large storage and memory for the large covariance matrix. On the other hand, randomized SVD (rSVD) \cite{halko2011finding,Halko_2011,Petros_2006,Sarlos} offers an efficient way to approximate the dominant singular components for high-dimensional outputs, thus allowing for a scalable architecture for modern “big data” applications.


\begin{figure*}[h]
\centering
    \includegraphics[width=.7\textwidth]{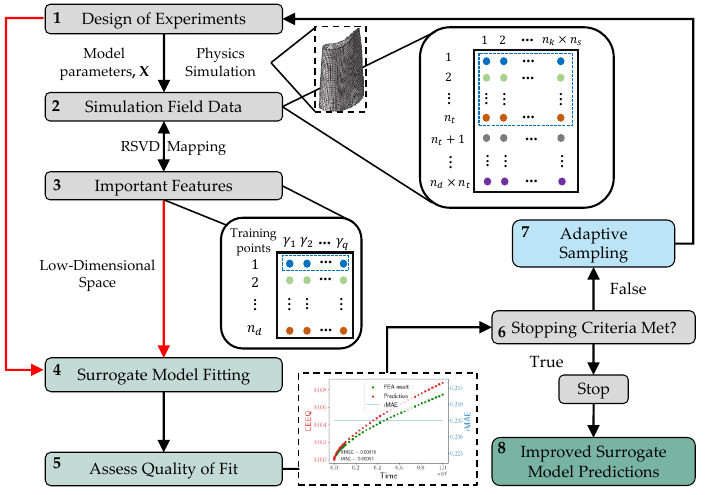}
  \caption{Schematic describing the proposed procedure for surrogate model construction with high-dimensional field data using rSVD}
  \label{fig:framework}
\end{figure*}

The quality of any surrogate model is critically influenced by the set of experiments or computer simulations of the original model used for training the surrogate model. Further, if the surrogate model is constructed in a low-dimensional space, in addition to the surrogate model error in the low-dimensional space, there is also reconstruction error in translating the surrogate model prediction to the high-dimensional original space. Thus, errors need to be quantified in a systematic and useful manner to assess the quality of the surrogate model. However, often the available resources prevent obtaining a large amount of training data when the original model is computationally expensive. Therefore, it is important to construct an accurate surrogate model with the fewest possible experiments or computer simulations. Design of experiments (DoE)~\cite{sacks1989design}, which refers to the selection of the input values at which to conduct these experiments, becomes crucial for sufficiently accurate model construction with minimum computational or experimental cost. One of the common DoE approaches is the space-filling approach that uses a distance metric to spread the samples evenly in the design space without considering any information about the physics of the problem.

In general, DoE methods can be classified as being non-adaptive or adaptive. The adaptive DoE techniques consider the physics of the system (as indicated by the previously sampled training points) while sampling new training points. Examples of non-adaptive DoE methods are the full factorial~\cite{box19612}, Latin hypercube~\cite{mckay2000comparison}, orthogonal array~\cite{owen1992orthogonal}, minimax and maximin-distance designs~\cite{johnson1990minimax}, where the focus is on coverage of the input space and not the physics of the input-output relationship. As a result, the non-adaptive DoE methods may under/oversample and waste computational resources. Whereas, in sequential adaptive sampling, the information obtained from the previous samples is used to populate new samples in regions of \emph{high interest}, e.g., where the underlying function is highly nonlinear or exhibits abrupt changes and the surrogate model accuracy is poor. Moreover, with adaptive sampling it is possible to stop the computationally expensive sampling process as soon as the surrogate model accuracy reaches an acceptable level.

Two components are essential in identifying additional sampling points to improve the surrogate model: (1) quantification of surrogate model error, and (2) decision criterion or learning function to identify the additional sampling points. Several approaches have been proposed in the literature to address both components. For example, cross-validation (CV) methods use estimates such as Leave-One-Out (LOO) CV error~\cite{jin2002sequential,li2010accumulative} and new training points are located in the region of the maximum value of the CV error. Adaptive approaches based on cross-validation variance (CVV) select new samples based on the predicted maximum value of the CVV among candidate future samples~\cite{kleijnen2004application,romero2006adaptive}. Hombal and Mahadevan \cite{hombal2011bias} proposed adaptively selecting training points that focus on minimizing the bias in the prediction. The LOO error-based Accumulative Error (ACE) approach~\cite{li2010accumulative} uses a weighted combination of LOO errors. Another method is based on mean squared error (MSE) \cite{jones2001taxonomy,jin2002sequential} which chooses the next training point based on the maximum value of the estimated mean squared error in the response predicted by a GP surrogate model. The Cross-Validation Voronoi (CVVor) is based on the combination of a cross-validation exploitation with a distance-based exploration, in which a Voronoi tessellation is employed to divide the whole input space into a set of Voronoi cells~\cite{xu2014robust}. The maximin scaled distance (MSD) method~\cite{jin2002sequential} uses a modification of maximin distance-based sampling that assigns weights to the important variables by using the available information. The Expected Improvement (EI) method uses geometry-based exploration and exploitation based on the variance of the prediction. The main goal of EI is to predict the global minimum value of the response accurately~\cite{jones1998efficient}. Another geometry-based exploitation method is Local Linear Approximation (LOLA)-Voronoi, which is a discontinuous adaptive sampling approach based on an exploitation feature with gradient estimation and exploration given by the volume of Voronoi tessellation cells~\cite{crombecq2011novel}. Adaptive methods using query-by-committee-based exploration such as Mixed Adaptive Sampling Algorithm (MASA)~\cite{eason2014adaptive} are also studied in the literature, where a new sample point is found by combining a local exploitation contribution based on Query by Committee (QBC) fluctuation and a global exploration based on distance.

Adaptive sequential sampling design (also referred to as active learning \cite{cohn1996active,yang2015active}) has also been investigated for reliability analysis, i.e., for the estimation of probability of failure; the focus of such studies has been accurate surrogate modeling of only the \emph{limit state}, often formulated as $g() = 0$, which is the boundary between regions of success and failure, instead of modeling the function $g() = 0$ over the entire domain. The efficient global reliability analysis (EGRA) \cite{bichon2008efficient} method extends the expected improvement (EI) \cite{jones1998efficient} mentioned above to define a new learning function Expected Feasibility Function (EFF), in the context of Gaussian process (GP) surrogate modeling of the limit state $g() = 0$. Other active learning methods for reliability analysis are AK-MCS \cite{ECHARD2011145} and AK-MCS + U \cite{PEIJUAN2017185} that adaptively select new training points for the GP model improvement. Based on the AK-MCS + U method, Hu and Mahadevan \cite{hu2016global} proposed an enhanced surrogate model-based reliability analysis method based on global sensitivity analysis to further improve the computational efficiency. All of the above-mentioned methods focus on approximating only the limit state $g() = 0$, whereas our concern in this paper is with a general prediction model, i.e., predicting the output accurately over the entire input space.

Overall, the limitations of the above-mentioned adaptive learning methods can be summarized as follows: (a) looking at a single limit state instead of the entire input space, (b) considering only single scalar output, and (c) not easily applicable to time-dependent multi-physics dynamic systems with high-dimensional spatio-temporal outputs. In order to address these limitations an adaptive surrogate modeling strategy is developed in this paper to predict the high-dimensional spatial and temporal output quantities of interest (QoIs).

The proposed approach as shown in Fig.~\ref{fig:framework} addresses two challenges: high-dimensionality of the output and adaptive selection of training runs for the surrogate model. For dimension reduction, the method of randomized singular value decomposition (rSVD)~\cite{halko2011finding} is used to identify the important features in the output space to give a lower dimensional representation of the original QoIs. These important features are then used to construct the surrogate model in the low dimensional space. Subsequently, the trained model is used to predict the QoIs in the original space. Error analysis is performed both in the lower dimensional latent space and the high-dimensional original space. This helps to quantify the contributions of surrogate model error and reconstruction error separately. This lays the foundation for adaptive improvement of the surrogate model. A novel approach that combines the ideas of exploration and exploitation for adaptive training point selection was developed in a manner that is applicable to time-dependent multi-physics dynamic problems with high-dimensional spatio-temporal outputs. 

The important contributions of this paper can be summarized as (1) Dimension reduction for very high-dimensional spatio-temporal outputs; (2) Surrogate model error quantification in two spaces; and (3) New sequential sampling technique for adaptive surrogate model improvement for multivariate spatio-temporal outputs.

The rest of this paper is organized as follows. A brief introduction to dimension reduction techniques is given in Section \ref{Sec:back}. The proposed method for adaptive surrogate model construction with high-dimensional spatio-temporal output is presented in Section \ref{Sec:Methodology}, which includes two aspects: error quantification and adaptive sampling strategy. Section~\ref{sec:proposedmethodcomparison} compares the proposed adaptive sampling technique with existing adaptive sampling algorithms, using several benchmark problems. Section \ref{Sec:Numerical example} demonstrates the application of the proposed method for gas turbine blade thermo-mechanical analysis (with spatio-temporal, multivariate output). Section \ref{Sec:Conclusion} provides concluding remarks and identifies future research needs.

\section{Background: Output Dimension Reduction}\label{Sec:back}

The model outputs considered in this paper vary over space and time, and are very high-dimensional. When there are many high-dimensional output QoIs at numerous spatial and temporal locations, it is not trivial to build surrogate models for every output at every location and time instant. Instead of building surrogate models directly in the original space, first the dimensionality of the high-dimensional response field can be reduced by mapping it to an uncorrelated latent space, and then a surrogate model can be constructed only for the important features in this latent space. This section discusses two techniques for dimension reduction: singular value decomposition (SVD) and randomized SVD (rSVD).

Singular value decomposition (SVD) is a generalized eigen-decomposition technique for describing a large amount of high-dimensional data by mapping to a low-dimensional latent space~\cite{chatterjee2000introduction}. SVD is applicable to non-square matrices and can be used to handle the spatial correlation of the response. Performing a basic SVD on a very large matrix is not only computationally expensive but also memory-intensive due to the need to store and invert very large matrices. Therefore, the randomized SVD (rSVD) algorithm, which requires less memory and avoids the high computational cost while not sacrificing accuracy, is pursued in this study to obtain a low-rank approximation of the large response matrix~\cite{halko2011finding}.

The randomized SVD (rSVD) method \cite{halko2011finding} maps the original data matrix to a small random subspace. The main computational operations in the rSVD-based approach used here are matrix-matrix multiplications, and QR decomposition and SVD on small matrices. More specifically, the rSVD involves five main components: (1) constructing a random matrix; (2) projection of the original data matrix to the random matrix; (3) QR decomposition of the resulting matrix; (4) multiplying the original data matrix with the resulting orthogonal basis; and (5) basic SVD on the resulting matrix.

In order to obtain a low-rank, say rank $r\ (r<<N)$, approximation of a matrix $\mathbf{X} \in \mathbb{F}^{M \times N}$,
\begin{equation}
\mathbf{X}\approx\mathbf{U}_r \mathbf{\Sigma}_r \mathbf{V}^{\prime}_r
\end{equation}
\noindent the target dimension, i.e., the first $r$ pairs of singular value and singular vectors, can be obtained with the following steps. First, a random projection matrix $\mathbf{P} \in \mathbb{R}^{N\times r}$ is used to get an orthonormal basis for the data matrix $\mathbf{X}$,
\begin{equation}
    \mathbf{Z} = \mathbf{X}\mathbf{P},
\end{equation}
\noindent where $\mathbf{Z} \in \mathbb{R}^{M\times r}$ is a much smaller matrix ($r<<N$) than $\mathbf{X}$ and approximate the column space of $\mathbf{X}$ with high probability due to the randomness of $\mathbf{P}$. Then, a QR decomposition, which is normally employed to compute the SVD, of $\mathbf{Z}$ can be obtained:
\begin{equation}
    \mathbf{Z} = \mathbf{Q}\mathbf{R}^{QR},
\end{equation}
\noindent where $\mathbf{Q} \in \mathbb{R}^{M\times r}$ and $\mathbf{R}^{QR} \in \mathbb{R}^{r\times r}$. Then, the data matrix $\mathbf{X}$ is projected onto the subspace $\mathbf{Q}$ and SVD is performed on the projection $\mathbf{Y} \in \mathbb{R}^{r\times N}$ to obtain $\mathbf{\Sigma}_r$ and $\mathbf{V}_r^T$: 
\begin{align}\label{eq: SVDY}
    \mathbf{Y} &= \mathbf{Q}^T\mathbf{X}, \nonumber \\
    \mathbf{Y} &= \mathbf{U}_Y\mathbf{\Sigma}_r\mathbf{V}_r^T.
\end{align}
\noindent The matrices $\mathbf{\Sigma}_r$ and $\mathbf{V}_r^T$ in Eq. \eqref{eq: SVDY} are the same for $\mathbf{X}$ since $\mathbf{Q}$ approximates the column space of $\mathbf{X}$ \cite{halko2011finding}. Thus, the $r$ left singular vectors $U_r \in \mathbb{R}^{M\times r}$ of $\mathbf{X}$ is
\begin{equation}
    \mathbf{U}_r = \mathbf{Q}\mathbf{U}_Y.
\end{equation}

A lower dimensional representation (dimension $r$) in place of the original data (dimension $N$) can be taken as
\begin{equation}
    \hat{\mathbf{X}}^{LD} = \mathbf{U}_r\mathbf{\Sigma}_r,
    \label{eq: SVDcoord}
\end{equation}

Randomized SVD (rSVD) is also a linear mapping like the basic SVD. Since $\mathbf{Z}$ is of a much smaller size compared to $\mathbf{X}$ and the basis $\mathbf{Q}$ is low rank, the QR decomposition in Step (3) and the basic SVD in Step (5) are not time consuming. This makes rSVD a powerful tool when dealing with a very large data matrix.

\section{Proposed Methodology}\label{Sec:Methodology}

The proposed methodology aims to efficiently construct a surrogate model for high-dimensional spatio-temporal response prediction. In this section, we first discuss dimension reduction to obtain a lower dimensional representation, \emph{features}, where we adapt the rSVD method for high-dimensional spatio-temporal output using a two-step mapping strategy. Next, we investigate multiple options for surrogate modeling and select the best surrogate model for a multivariate output. This is achieved by performing a cross-validation analysis for each multivariate surrogate model constructed in a low-dimensional space for the important features. Following this, we develop the adaptive sampling strategy for identifying additional training points to improve the surrogate model, by combining exploitation and exploration. Thus the proposed overall methodology consists of four steps:
\begin{enumerate}
    \item Dimension reduction (two-step mapping)
    \item Surrogate model construction
    \item Surrogate model error quantification (through cross-validation) 
    \item Adaptive training point selection
\end{enumerate}
The following subsections describe these steps in detail. 

\subsection{Dimension Reduction}\label{sec:method_dimred}
Given $n_s$ data points (i.e., total number of nodes) over the spatial domain $\mathbf{\Omega}$ for $n_t$ time domain realizations and for $n_d$ realizations of design domain (i.e., training points), a data matrix for the $k$-th output can be written as
\begin{footnotesize}
\begin{align}
    &\mathbf{D_k}=[\mathbf{D_k}(t_i,\theta_1), \mathbf{D_k}(t_i,\theta_2), \dots, \mathbf{D_k}(t_i,\theta_{n_d})]^T \notag \\ 
    &= \begin{bmatrix} 
    D_k(t_i,s_1, \theta_1) & D_k(t_i, s_1, \theta_2) & \dots & D_k(t_i,s_1,\theta_{n_d}) \\
    D_k(t_i,s_2, \theta_1) & D_k(t_i,s_2, \theta_2) & \dots & D_k(t_i,s_2,\theta_{n_d}) \\
    \vdots & \vdots & \ddots & \vdots \\
    D_k(t_i,s_{n_s},\theta_1) & D_k(t_i,s_{n_s},\theta_2) & \dots & D_k(t_i,s_{n_s},\theta_{n_d}) 
    \end{bmatrix}^T
\end{align}
\end{footnotesize}
\noindent where $\mathbf{D_k}(t_i,\theta_j)=[D_k(t_i,s_1,\theta_j), D_k(t_i,s_2, \theta_j),\allowbreak \dots,D_k(t_i,s_{n_s},\theta_j)]$ is the $i$-th temporal location of the $j$-th realization for the $k$-th field response, $t_i$ represents the $i$-th temporal location and $\mathbf{\theta}$ stands for different realizations in the design domain. 

The response $\mathbf{S}$ of the original physics-based model can be collected at the training points as follows:
\begin{equation}
    \mathbf{S} = [\mathbf{D_1},\mathbf{D_2},\dots,\mathbf{D_{n_k}}]^T
\end{equation}
\noindent where $n_k$ is the total number of physics model outputs.

This large amount of high-dimensional response can be mapped to a low-rank approximation by using rSVD as $\mathbf{S}=\mathbf{UMV}^T$, where $\mathbf{U}$ is a $(n_t \times n_d) \times n_s$ matrix, $\mathbf{V}$ is a $n_s \times n_s$ orthogonal matrix and $\mathbf{M}$ is a $n_s \times n_s$ rectangular diagonal matrix with non-negative real numbers $\mathbf{\lambda}=[\lambda_1,\lambda_2,\dots,\lambda_r$], $r=\text{min}((n_t \times n_d),n_s)$ in the diagonal. The diagonal elements $\mathbf{\lambda}$ of $\mathbf{M}$ are called singular values and are arranged in descending order. The number of important features $q (q\leq r)$ to be used to represent the abstract features of the original data matrix $\mathbf{S}$ are determined based on the magnitudes of $\mathbf{\lambda}$. By using the first $q$ largest singular values, $\mathbf{S}$ can be reconstructed as $\mathbf{\Tilde{S}}$:
\begin{align}
 \mathbf{S}(t_i,\theta_j)^T&\approx\mathbf{\tilde{S}}(t_i,\theta_j)=\sum_{p=1}^q \xi_p(t_i,\theta_j)\mathbf{V}_p,\\ \notag
 & \forall i=1,2,\dots,n_t;\  j=1,2,\dots,n_s,   
\end{align}
where $\mathbf{S}(t_i,\theta_j)^T$ is the $(i\times j)$-th row of $\mathbf{S}$, $\xi_p(t_i,\theta_j)$ is the element of $\xi=\mathbf{UM}$ at $(i\times j)$-th row and $p$-th column, and $\mathbf{V}_p$ is the $p$-th important feature vector used to approximate $\mathbf{S}$.

The rSVD method described in Section \ref{Sec:back} is extended in this work for dimension reduction. The following additional steps are proposed to adapt this method for a high-dimensional spatio-temporal output, in a manner that facilitates surrogate model construction. 

A two-step dimension reduction approach is employed to identify the important features (principal components) in the low-dimensional output space to give a lower dimensional representation of $\mathbf{S}$. The dimension of the output data matrix $\mathbf{S} \in \mathbb{R}^{(n_t\times n_d)\times n_s}$ has a large number of features and it has more features (columns) than observations (rows). Thus, in the first step of the two-step approach the original features of the output data matrix can be reduced using rSVD to a smaller subset of features that are most relevant to the prediction problem. The result is matrix $\mathbf{I}^{(1)} \in \mathbb{R}^{(n_t\times n_d)\times n_1}$ with a lower rank $n_1$ that is said to approximate the original spatio-temporal output $\mathbf{S}$. Let us denote the truncated singular value matrix, eigenvector matrix, and orthogonal matrix obtained using $n_1$ important features as $\mathbf{U}^{(1)} \in \mathbb{R}^{(n_t\times n_d)\times n_1}$, $\mathbf{\Sigma}^{(1)} \in \mathbb{R}^{n_1\times n_1}$, $\mathbf{V}^{(1)} \in \mathbb{R}^{n_1\times n_1}$. The compressed form of $\mathbf{I}^{(1)}$ is $\mathbf{I}^{(1)}\approx \mathbf{U}^{(1)} \mathbf{\Sigma}^{(1)} (\mathbf{V}^{(1)})^T$.

The resulting matrix $\mathbf{I}^{(1)}$ still has time-dependency. The rows $\mathbf{I}^{(1)}$ include the temporal information and the realizations of design domain. The columns of $\mathbf{I}^{(1)}$ represent the time series features. It was not feasible to perform a single rSVD and reduce dimensions for all quantities of interest both in time and space at the same time. We preferred to remove this temporal correlation in the second step of the two-step approach for two reasons. First, surrogate model construction was simpler for time-independent features. Second, the accuracy of the surrogate model increased as we removed the temporal variation with the proposed two-step dimension reduction approach. 

When rSVD is performed twice on the transformed version of matrix $\mathbf{I}^{(1)}$, i.e., $\mathbf{I}^{(1)'} \in \mathbb{R}^{n_d\times (n_1\times n_t)}$, where the rows represent the realizations of design domain, and the columns represent the time series features for each design, we obtain $\mathbf{I}^{(2)}$. The columns of this low-rank matrix $\mathbf{I}^{(2)}$ would represent the important features (i.e., the number of features $n_2$ that are not time series to build the surrogate model) and the rows represent the FEM simulations performed with different realizations of the design domain. Thus, we obtain $\mathbf{I}^{(2)}\approx \mathbf{U}^{(2)} \mathbf{\Sigma}^{(2)} (\mathbf{V}^{(2)})^T$, where $\mathbf{U}^{(2)}$ $\in$ $\mathbb{R}^{n_d\times n_2}$, $\mathbf{\Sigma}^{(2)}$ $\in$ $\mathbb{R}^{n_2\times n_2}$, $\mathbf{V}^{(2)}$ $\in$ $\mathbb{R}^{n_2\times (n_1\times n_t)}$.

The accuracy of the above two-step dimension reduction needs to be evaluated. To do this, the low-rank approximated matrix $\mathbf{I}^{(2)}$ in the second feature space is mapped to the original space (i.e., first it is mapped to the first feature space and then mapped to the original space). The reconstruction accuracy is quantified using root mean square error (RMSE). The RMSE value decreases with the number of important features used to approximate the original matrix $\mathbf{S}$. The number of important features is chosen based on the reconstruction accuracy and the percentage of variance explained by the top few singular vectors and singular value pairs. The proposed two-step dimension reduction approach is used in the numerical example in Section \ref{Sec:Numerical example}. Next, the surrogate model construction described in Section \ref{sec:surrmodel} is performed in the second low-dimensional space using the important features $\boldsymbol{\xi}^{(2)}=\mathbf{U}^{(2)}\mathbf{\Sigma}^{(2)}$.

\subsection{Surrogate Model Construction}\label{sec:surrmodel}
Any of the available surrogate modeling techniques mentioned earlier (e.g., PCE, GP, SVR, DNN etc.) can be used to construct the surrogate model. In addition to these techniques, other prominent machine learning algorithms based on boosting (which is one of ensemble learning algorithms) are also available for improving the performance of a simple machine learning model. This paper explores several of them such as Light Gradient Boosting Machine (lightGBM) \cite{NIPS2017_6449f44a}, Extreme Gradient Boosting (XGBoost) \cite{xgboost}, Categorical Boosting (CatBoost) \cite{dorogush2018catboost}, and random forest (RF) \cite{hastie2009elements}. Gradient Boosting Decision Tree (GBDT) is a relatively new decision tree-based ensemble learner. LightGBM, XGBoost, and CatBoost are different variations of gradient boosting methods. The main difference between these techniques is how they build the decision tree.

A tree-based regression technique known as extremely randomized tree regressor or simply Extra-Trees regressor~\cite{geurts2006extremely} is found to give the best performance in the numerical example (Section \ref{Sec:Numerical example}), based on comparing the k-fold cross validation error results for various techniques. Tree-based ensemble methods randomly construct more than one decision tree to achieve an increase in the generalization performance. For example, random forest (RF) regression~\cite{hastie2009elements} trains the trees with bootstrap samples for each candidate split based on randomly selected subset of the features. The main idea behind Extra-Trees is to randomly create a number of different trees with randomly chosen features. The randomization helps to achieve a greater reduction in the variance of the model prediction, compared with other ensemble methods like RF~\cite{hastie2009elements}. The main differences between the Extra-Trees algorithm and RF regression are:~(a) Extra-Trees randomly splits nodes using a random subset of the features selected at every node, rather than the best split used in RF; and~(b) RF applies the bagging procedure to iteratively generate sub-training sets (bootstrap samples) with replacements, while Extra-Trees uses all the training samples to construct each tree with varying numbers of parameters. 

In this work, the surrogate model is not only built for the design stage, but also for future health management of the component. Depending on how the component is used different quantities could be critical at different times. Thus, the surrogate model is envisioned to handle such changing requirements.

\subsection{Surrogate Model Error Quantification}\label{sec:LOOCV}
The surrogate model prediction error is used to inform about areas with the highest error. The Leave-One-Out Cross-Validation (LOOCV) method is a special case of the $k$-fold cross-validation (CV), with $k=N$. For each observation $i\in [1,N]$, a separate surrogate model $\mathcal{M}_{-i}$ is trained on $N-1$ observations consisting of the reduced set $\mathcal{D}_{-i}=\mathcal{D}\backslash(\mathbf{x}^i,\mathbf{y}^i)$. The surrogate model accuracy is then evaluated on the test point $\mathbf{x}^i$. The spatio-temporal physical quantities are predicted for the test data using the surrogate model trained in the latent space, where high-dimensional spatio-temporal outputs are projected to. The surrogate model outputs in the feature space are then mapped to the original space (reconstruction) to evaluate the surrogate model accuracy in the original space. 

Different error metrics are suitable in different situations. For example, an engineer may be more interested in the relative error of creep damage response in the regions where strong creep behavior is expected since a small amount of change can lead to detrimental effects. Whereas the engineer is less interested in the relative error of the predicted response in regions where weak creep behavior (i.e., physical quantities are close to zero) is dominant. Thus, for this purpose, two error metrics are calculated using LOOCV, namely mean absolute error (MAE) and relative mean absolute error (rMAE). The MAE metric is defined as follows:
\begin{equation}\label{eq:mae}
    \text{MAE}_j = \text{mean(AE$_j$)} = \frac{1}{n_{time}}\sum_i^{n_{time}} \lvert S_{i,j}-\hat{S}_{i,j}\rvert
\end{equation}
where the subscript $j$ refers to the $j$th FEM node (e.g., MAE$_j$ is the mean absolute error of $j$th node, where the mean is taken across time for each spatial location). Using MAE will allow the engineers to choose a different threshold that has a physical meaning for each quantity of interest. The main challenge of the MAE metric is the identification of thresholds. Moreover, the MAE metric is scale-dependent. In order to address such limitations, the relative mean absolute error (rMAE) metric that is scale independent and less sensitive to outliers is investigated. The rMAE metric is defined as follows:
\begin{equation}\label{eq:rmae}
    \text{rMAE} = \frac{\text{MAE}}{\text{mean}\lvert\mathbf{S}-\mathbf{\bar{S}}\rvert}
\end{equation}
where $\mathbf{\bar{S}}$ is the mean of the time series output at each spatial location. An advantage of this method is its interpretability. For example, rMAE measures the possible improvement of the proposed model relative to the benchmark model (i.e., the denominator of Eq. \eqref{eq:rmae}). When rMAE is less than 1, the proposed model is better than the benchmark model, and when the value of this error metric is greater than 1, then the proposed model is worse than the benchmark model. The only circumstance under which rMAE would be infinite or undefined is when \emph{all} historical observations are equal (i.e., static problem). 

For each node in the FEM model, the error metrics are evaluated based on the LOOCV predictions. Note that the surrogate model predicts features in the low-dimensional latent space, which are then mapped to the original space to evaluate the error metrics for the spatio-temporal physical outputs. Then, PNMAE, the percentage of nodes having MAE values greater than some threshold value, and PNrMAE, the percentage of nodes having rMAE values greater than one, are calculated for the current training points using LOOCV. However, there is no way to assess these metrics for points not in the current design, i.e., unobserved points (or candidate new points, $\mathbf{x}^c$). Therefore, additional surrogate models are built to learn the relationship between the training points and the corresponding LOOCV prediction error metrics in the original space. In this study, Gaussian process (GP) surrogate models are used (allowing the use of Expected improvement (EI) \cite{jones1998efficient}) to predict $e_{PNMAE}$ and $e_{PNrMAE}$ for unobserved points. These GP models are developed using the initial training points. As new points are added to the design $\mathcal{D}$, the training data used to train these GP models are updated.

\subsection{Proposed Adaptive Sample Selection Method}\label{sec:EICV}
First, the approach generates an initial design. The initial design is chosen using a space-filling criterion with a given initial number of points. In this work, we use maximin Latin hypercube sampling (LHS) \cite{morris1995exploratory,jin2003efficient}, though any other space-filling metric can be used, to generate initial training points that cover the input space uniformly. Based on this initial design, the LOOCV error $e_{\rm LOOCV}$ is calculated for each point in the design space by building a surrogate model each time leaving one point out. A secondary GP surrogate model is constructed for $e_{\rm LOOCV}$ to predict $e_{\rm LOOCV}$ for candidate points for additional samples.

A straightforward approach is to simply use the LOOCV error for adaptive sampling, i.e., exploit regions with high prediction error to select new training points \cite{viana2021surrogate}. However, a purely exploitation-based approach can result in clustered samples in the input space. In order to avoid this problem, an exploration strategy based on a space-filling criterion can be added to spread the new samples over the entire input domain while exploiting the regions of interest.

Thus two strategies, namely exploitation and exploration, are combined here for surrogate model improvement through adaptive selection of additional training points. Exploration aims to discover regions of the input space not covered in previous training points in order to obtain knowledge of the response over the entire design space; thus exploration does not use the simulation outputs for the previous training points. Whereas exploitation uses the output information gained from previous training points to identify high-interest subregions to generate new samples in the vicinity of these regions. These subregions can be associated with large prediction error, significant non-linear behavior, discontinuity, etc. New samples can be added in the region of interest depending on the aim of the analysis to build a surrogate model that predicts the response accurately in the exploited high-interest regions. Instead of considering these strategies individually, it can be beneficial to consider them together in hybrid adaptive learning to leverage both of their strengths, such that exploration adds points from previously unexplored regions (geometry-based) and exploitation adds points in regions of interest pertaining to surrogate model accuracy (physics-based).

\subsubsection{Space-Filling Criterion}\label{sec:sfc}
A space-filling criterion is used to avoid the clustering of samples that could occur in a purely exploitation-based approach and ensure uniform coverage of the design space. The space-filling criterion used in this work is based on Euclidean distance, specifically the maximin distance, wherein the minimum non-zero distance $d_{min}$ of a candidate point from all other points in the current set of training points is computed. A candidate point ($\mathbf{x}^c$) with the maximum of these minimum distances is selected as the new training point from $\mathbf{x}^{*}$. Thus, the exploration strategy is given by:
\begin{align}
    \mathbf{x}^{c}&=\arg \max_{\mathbf{x}^{*} \in \mathbb{X}} \left(d_{min}\right) \nonumber \\ 
    d_{min}&=\min_{\substack{\mathbf{x}^i \in \mathcal{X} \\ \mathbf{x}^{*} \in \mathbb{X} }} ||\mathbf{x}^i-\mathbf{x}^{*}||
\end{align}

\subsubsection{Expected Improvement}\label{sec:EI}
In order to overcome the \emph{overexploitation} problem, which could cause new samples to be clustered in regions with large mean LOOCV prediction error, a well-known approach is to use the \emph{Expected improvement} (EI) \cite{jones1998efficient} function, which helps to trade-off between exploitation and exploration. It can be expressed as
\begin{equation}
\resizebox{0.5\hsize}{!}{%
        $EI(\mathbf{x}) =
        \begin{cases}
            \left(\mathbf{y}^{*} - \mu(\mathbf{x})\right)\Phi(u) + \sigma(\mathbf{x})\phi(u), & \text{if $\sigma(\mathbf{x}) > 0$}.\\
            0, & \text{if $\sigma(\mathbf{x})=0$}.
        \end{cases}$
        }
\end{equation}
where $u=\left(\mathbf{y}^{*}-\mu(\mathbf{x})\right)/\sigma(\mathbf{x})$, $\mathbf{y}^{*}$ being the current best solution chosen from among the true function values at the training points and $\phi(\dot)$ and $\Phi(\dot)$ represent the PDF and CDF of the standard normal distribution, respectively. EI is a non-negative, parameter-free function and is zero at the training points.

\subsubsection{Proposed Learning Function}\label{sec:learningfunc}
Combining the EI function defined in Section \ref{sec:EI} and the space-filling criterion described in Section \ref{sec:sfc}, a point with the largest expected improvement $EI_{\rm PNMAE}$ (i.e., expected improvement based on PNMAE) and $EI_{\rm PNrMAE}$ (i.e., expected improvement based on PNrMAE) and the maximum $d_{min}$ is selected as the new sample point. Thus, a learning function that facilitates the active learning process is proposed here as:
\begin{equation}\label{eq:learningfunc}
\resizebox{0.5\hsize}{!}{%
$\mathcal{L}(\mathbf{x})= (\beta_1 EI_{\rm PNMAE} +\beta_2 EI_{\rm PNrMAE})^{\alpha} \times (d_{min}(\mathbf{x}))^{\gamma}$
}
\end{equation}
where $\beta_1, \beta_2 \in [0,1]$ control the relative contributions of $EI_{\rm PNMAE}$ and $EI_{\rm PNrMAE}$ respectively, and $\alpha,\gamma \in [0,1]$ are used to control the trade-off between exploitation and exploration respectively. (Other mathematical formulations of the learning function using these metrics are also possible). In this formulation of the learning function, EI provides both local exploitation and local exploration based on the mean and variance of the prediction error, whereas the space-filling criterion only provides geometry-based global exploration. The surrogate model variance quantifies the closeness between the new training points and all existing training points. Reduction of such variance may only be a local exploration depending on the error metrics used. The error metrics used in the engineering application are dependent on the actual values of the QoIs as discussed in Section~\ref{sec:adap_app}. Thus, here we used the space-filling criterion that focuses on the sample distances in the input space and guarantees global exploration. Note that $\mathcal{L}$ provides a trade-off between the space-filling criterion, which avoids clustering, and the largest estimation of prediction error. 

The two extreme cases, $\alpha=0$ and $\alpha=1$, respectively denote that exploitation is not considered and exploitation is considered. Exploration of the input space is controlled by $\gamma$. When $\gamma=0$, the minimum Euclidean-distance is not considered (i.e., no exploration), whereas the larger the $\gamma$ value is, the higher the weight of the exploration term. 

The parameters of the learning function can be estimated by using an algorithm similar to the Expectation-Maximization (EM) algorithm~\cite{dempster1977maximum}. We have 2 sets of parameters. First, we fix the first set \{$\beta_1,\beta_2$\} and compute the second set \{$\alpha,\gamma$\} in one step. Then, we compute the first set based on the computed values of the second set in the previous step. These two steps are repeated until the change in the parameter estimates between consecutive iterations does not exceed a desired threshold.

The new sample point $\mathbf{x}^c$ that results in the maximum learning function value $\mathcal{L}_{max}$ is identified as follows:
\begin{align}
\mathcal{L}_{max} &= \smash{\displaystyle\max_{\mathbf{x}^{*} \in \mathbb{X}}} \left[\mathcal{L}(\mathbf{x}^{*}) \right] \nonumber \\
\mathbf{x}^c &= \arg \mathcal{L}_{max}.
\end{align}

Note that the evaluations of candidate points using the proposed method are computationally inexpensive since these evaluations are based on the current surrogate model, not the original expensive physics model.

\subsubsection{Stopping Criteria}
The computational resource limit (i.e., number of training points, $N$), and the LOOCV error can be used as the stopping criteria. The latter criterion is used to evaluate the surrogate model performance on the test set. The surrogate model accuracy is evaluated as new samples are added to the design based on the proposed adaptive sampling approach. If the LOOCV error metrics $e_{\rm PNMAE}$ and $e_{\rm PNrMAE}$ achieve an acceptable value, then stop adding new points. In this work, both the computational resource limit and surrogate model accuracy are used as the stopping criteria. The detailed sequence of steps of the proposed adaptive learning approach is outlined in Algorithm~\ref{algorithm}.

\begin{algorithm*}[h]
    \small
  \caption{Identifying new training points for adaptive improvement of the surrogate model}\label{algorithm}
  \textbf{Input}: $\mathcal{D}$\\
  \textbf{Output}: $\mathbf{x}^c$\
  \begin{algorithmic}[1]
  
  \State Generate an initial design using maximin LHS
    \Procedure{Output Dimension Reduction}{}
    \State Construct the data matrix, $\mathbf{S}$
    \State Perform a two-level rSVD on the matrix $\mathbf{S}$ to obtain important features
    \EndProcedure
    \\
    \Procedure{Leave-One-Out Cross-Validation (LOOCV)}{}
    \State Split the design set $\mathcal{D}$ into $k$ disjunct sets $\mathcal{D}_i$, $i=1,\dots,k$, with $k=N$ (LOO)
        \For{$i= 1\ \text{to}\ N$}
            \State{Train an Extremely Randomized Trees machine learning model $\mathcal{M}_{-i}$ with $N-1$ features consisting of the reduced set $\mathcal{D}_{-i}=\mathcal{D}\backslash(\mathbf{x}^i,\mathbf{y}^i)$.}
            \State Calculate $e_{\rm PNMAE}$ and $e_{\rm PNrMAE}$ for the current training points in the reduced set $\mathcal{D}_{-i}$:
            \State $e_{\rm PNMAE}=$ \% of nodes having MAE $\ge$ 2.5e-4
                \State\quad $\text{MAE}_j = \text{mean(AE$_j$)} = \frac{1}{n_{time}}\sum_i^{n_{time}} \lvert S_{i,j}-\hat{S}_{i,j}\rvert$
            \State $e_{\rm PNrMAE}=$ \% of nodes having rMAE $\ge$ 1
                \State\quad $\text{rMAE} = \frac{\text{MAE}}{\text{mean}\lvert\mathbf{S}-\mathbf{\bar{S}}\rvert}$
        \EndFor
    \EndProcedure
    
    \\
    \Procedure{Adaptive Learning}{}
    \While{Average PNrMAE $\le$ 15\% or $N\ge 66$}
    \State Construct GP models for $e_{\rm PNMAE}$ and $e_{\rm PNrMAE}$ (LOOCV errors) to make predictions $\hat{e}_{\rm PNMAE}$ and $\hat{e}_{\rm PNrMAE}$ for candidate points
    \State Define the learning function:
    \State $\mathcal{L}= \left[(\beta_1 EI_{\rm PNMAE} +\beta_2 EI_{\rm PNrMAE})^{\alpha} \times (d_{min}(\mathbf{x}^{*}))^{\gamma} \right]$
    \State Sample a large set of random points from the design space and evaluate $\mathcal{L}$ for these points
    \State Identify the largest learning function value; $\mathcal{L}_{max}= \smash{\displaystyle\max_{\mathbf{x}^{*} \in \mathbb{X}}}\ \mathcal{L}(\mathbf{x}^{*})$
    \State Choose the new training point $\mathbf{x}^c$ that results in $\mathcal{L}_{max}$, i.e., $\mathbf{x}^c= \arg \mathcal{L}_{max}$.
    \EndWhile
    \EndProcedure
    
  \end{algorithmic}
\end{algorithm*}

\subsection{Summary of Methodology} 

The proposed methodology for adaptive surrogate modeling with high-dimensional spatio-temporal output as shown in Fig.~\ref{fig:framework} consists of the following steps:
\begin{enumerate}
    \item Construct the output data matrix $\mathbf{S}$ based on the available simulations where each row represents a time instant corresponding to one FEM run, and the columns represent multiple spatial locations of multiple output quantities for that FEM run;
    \item Perform the two-step dimension reduction approach explained in Section \ref{sec:method_dimred} to find the lower dimensional representation $\mathbf{I}^{(2)}$;
    \item Build a \emph{single} surrogate model for all features vs. process inputs in the feature space;
    \item For the test set points, project the surrogate model prediction in the feature space back to the original space to obtain the predictions in the original space, $\mathbf{\hat{S}}$;
    \item Use cross-validation to evaluate the surrogate model accuracy in the original space;
    \item If the accuracy of the surrogate model is not acceptable, then use the proposed adaptive sequential sampling based on the learning function given in Section \ref{sec:learningfunc} to propose additional training points for the surrogate model;
    \item Repeat steps 1 to 6 until the model accuracy reaches an acceptable value or until the computational resource limit reached.
\end{enumerate}

\section{Evaluation of Proposed Approach using Benchmark Problems}\label{sec:proposedmethodcomparison}

In this section, the proposed adaptive sampling technique is compared to several existing techniques described in Section \ref{sec:Intro} (i.e., EI, CVV, LOLA, MSD, MASA, CVVor). To provide a fair comparison, the performance of each sampling technique in accurately capturing seven benchmark test functions of different complexity is analyzed (see Appendix~\ref{app:1}). Since the $R^2$ values for all cases were close to each other, normalized root mean square error (NRMSE = $\text{RMSE}/(y_{max}-y_{min})$, where RMSE = $\sqrt{\sum_{i=1}^{n} (y_{i}- \hat{y}_{i})^2/n}$) is used to be able to distinguish them from each other. NRMSE  is first computed for each fold of LOOCV, followed by the computation of the average LOOCV error, by taking the average across all the LOOCV folds. This average LOOCV error is used to assess the performance of all the adaptive sampling techniques.

\begin{figure*}[h]
\centering
    \includegraphics[width=.8\textwidth]{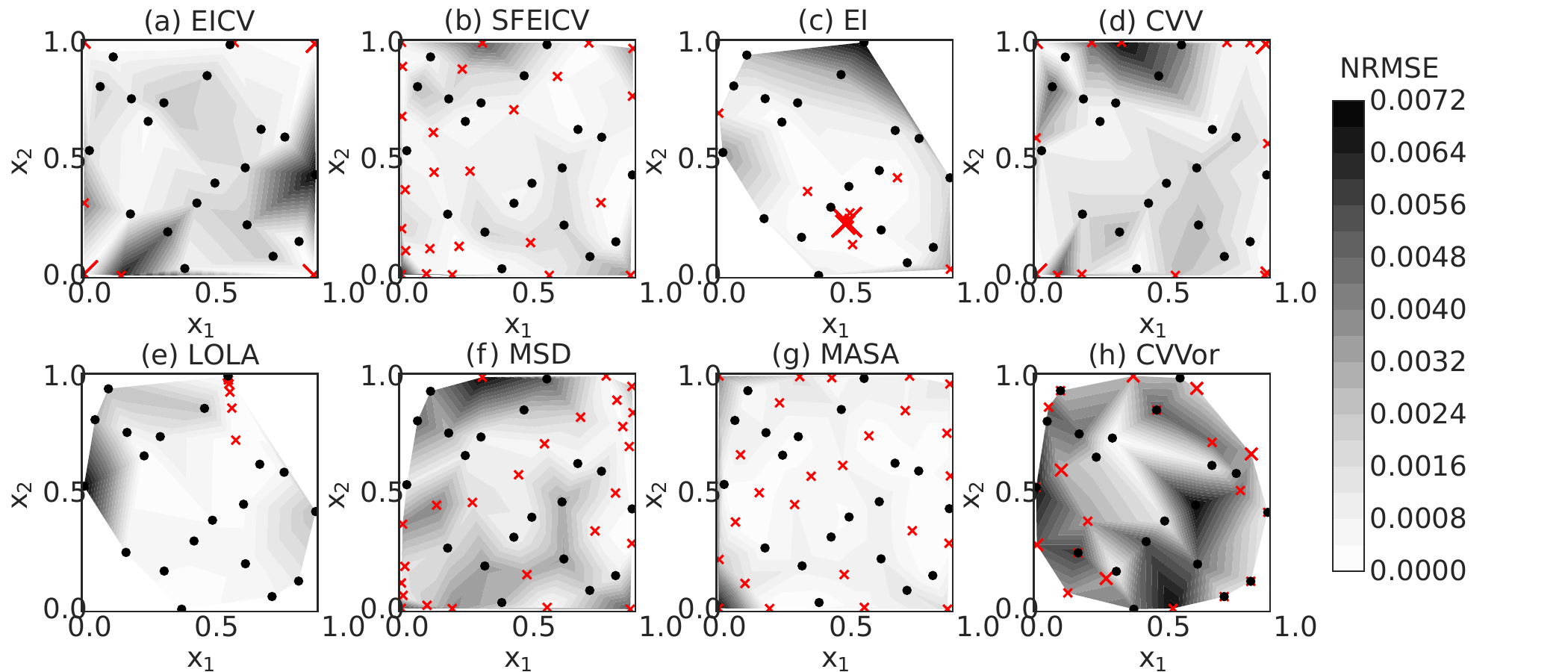}
  \caption{Adaptively selected training point locations for the Branin's function and the final surrogate model accuracy in terms of NRMSE are plotted in contours (Black dots: initial 20 samples, Red dots: additional 25 samples)}
  \label{fig:contmap_branin}
\end{figure*}

\begin{figure*}[h]
\centering
    \includegraphics[width=.8\textwidth]{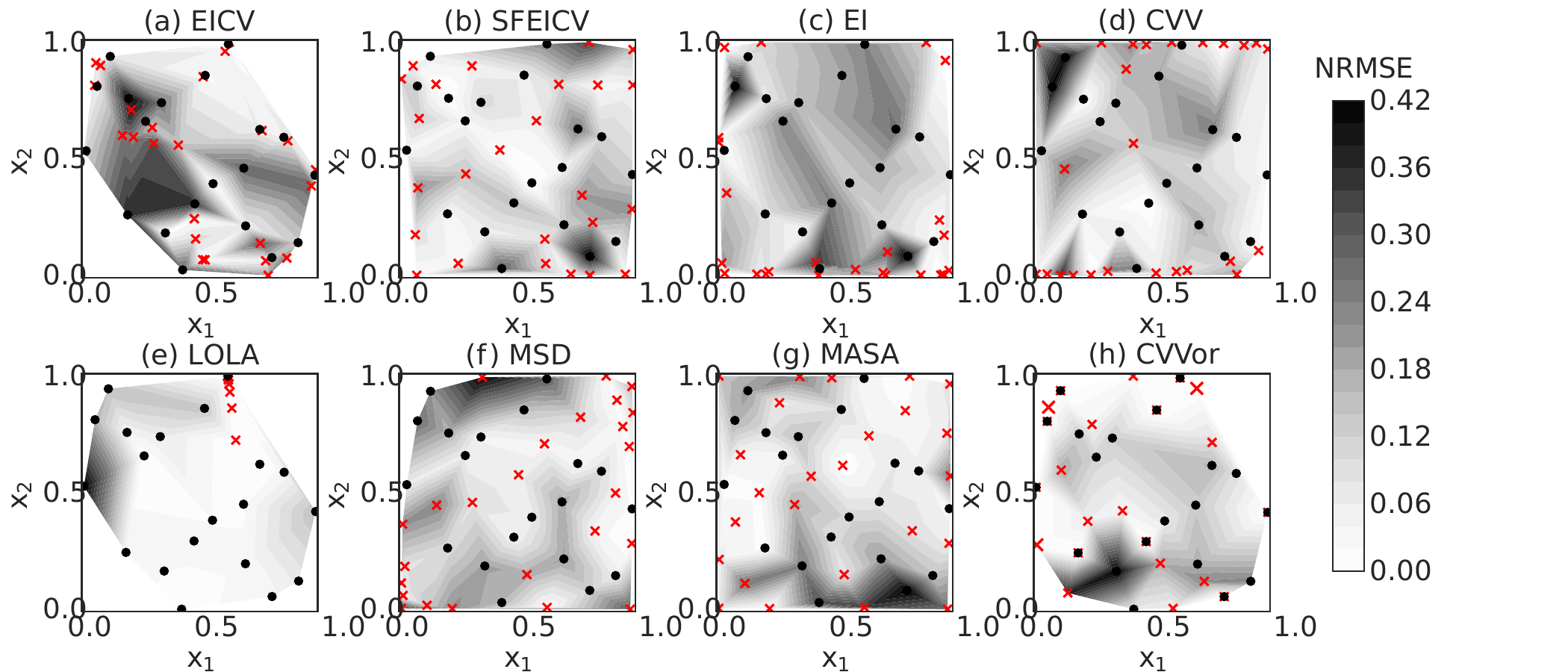}
  \caption{Adaptively selected training point locations for the Schwefel function and the final surrogate model accuracy in terms of NRMSE are plotted in contours (Black dots: initial 20 samples, Red dots: additional 25 samples)}
  \label{fig:contmap_sch}
\end{figure*}

Adaptive sampling techniques with higher exploitation component are strongly dependent on the size of the initial sample. A few empirical formulas have been proposed in the literature for choosing the initial sample size \cite{jones1998efficient,liu2016adaptive}. In this work, the number of samples included in the initial design ($m$) is chosen based on the rule $m=10n$ \cite{jones1998efficient}, where $n$ is the dimension of the input space. For each two-dimensional test function in Appendix~\ref{app:1}, an initial training set consisting of 20 points is generated using maximin LHS and used to start all the adaptive sampling techniques. The positions of the 20 initial sample points are shown by black dots in Figs.~\ref{fig:contmap_branin} and~\ref{fig:contmap_sch}. In order to have a fair comparison between the proposed method and the existing adaptive sampling techniques, a Gaussian process (GP) regression model is built with the exact same properties in each fold of LOOCV-based techniques. Then the number of training points is increased up to 45 by adding a new point in each step for each adaptive sampling technique, following each technique’s procedure.

\begin{figure*}[h]
\centering
    \includegraphics[width=.8\textwidth]{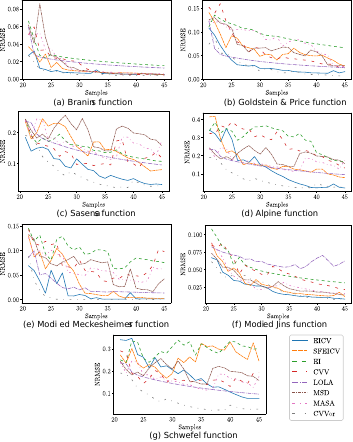}
  \caption{Evolution of the NRMSE from the initial surrogate model (20 samples) to the final surrogate model (45 samples) using different adaptive sampling techniques, for the seven benchmark functions in Appendix~\ref{app:1}}
  \label{fig:NMRSE}
\end{figure*}

The NRMSE results for different sampling techniques for 45 total samples are shown in Fig.~\ref{fig:NMRSE} for different benchmark functions. For illustration purposes, the adaptively selected samples are highlighted by red dots (larger points indicate a sample very close to a neighboring point) and the corresponding surrogate model accuracy in terms of NRMSE are shown as contours in Figs.~\ref{fig:contmap_branin} and~\ref{fig:contmap_sch} for the Branin and Schwefel functions in Appendix~\ref{app:1} respectively, for each sampling technique.

The Expected Improvement Cross-Validation (EICV) and Space-Filling Expected Improvement Cross-Validation (SFEICV) techniques are special cases of the proposed method; these two methods are compared against other adaptive improvement methods in this section. In EICV, $\gamma=0$, therefore the weighted minimum Euclidean distance is not considered (i.e., no additional exploration other than the exploration already inherent in EI). Whereas in SFEICV both $\theta$ and $\gamma$ equal 1, thus the weight of exploration is higher. (Note that two separate GP models are constructed for EICV and SFEICV, where the first GP model is to predict the original output QoI, and the second GP model is to predict the LOOCV error as described in Section \ref{sec:EICV}).

An issue with some techniques with the exploitation component is that they may focus on insignificant features with respect to the problem of interest. When this is the case, the technique cannot capture the proper behavior of the entire function and its performance becomes highly dependent on the initial dataset. In most complex engineering problems, the initial dataset is limited, thus it is important to select an adaptive technique such as the proposed EICV that does not focus on insignificant features of the problem and exploits the regions of interest. The gradient estimation approach used in Local Linear Approximation (LOLA)-Voronoi gets increasingly complex as the problems become high-dimensional and complex. The performance of LOLA is also not as good as EICV (see Figs.~\ref{fig:NMRSE}(a)-(g)), which is a more robust and computationally efficient technique for high-dimensional problems. It should be noted that all adaptively selected samples (i.e., red dots in Figs. \ref{fig:contmap_branin} and \ref{fig:contmap_sch}) using the LOLA technique are in regions where the initial surrogate model error was already low. This behavior of the LOLA technique can be observed in Fig.~\ref{fig:contmap_branin}(e). The computational complexity of Mixed Adaptive Sampling Algorithm (MASA) depends on the number of committee members (e.g., in the context of GP models, committee members could be autocorrelation functions). In these test problems, three GP models with different autocorrelation functions (Mat\'ern, squared exponential kernel with a different correlation length for each coordinate, and squared exponential kernel with same correlation length in all coordinates) are considered as committee surrogates (e.g., GP models based on different autocorrelation functions). MASA shows a higher emphasis on exploration (see Figs. \ref{fig:contmap_branin}(g) and \ref{fig:contmap_sch}(g)) and is not as dependent on the initial sample size as some other methods such as EI and MSD. The Maximin scaled distance (MSD) method performs poorly because the method is not capable of exploiting as much as the other methods (as seen in Figs.~\ref{fig:NMRSE},~\ref{fig:contmap_branin} and~\ref{fig:contmap_sch}). SFEICV is almost as good as MASA and MSD at exploring the design space (see Figs. \ref{fig:contmap_branin}(b) and \ref{fig:contmap_sch}(b)), and it can also exploit the regions where there is high prediction error due to local non-linearity (note the additional samples selected on the bottom left corner in Fig.~\ref{fig:contmap_branin}(b)).

It should be noted that Cross-Validation Voronoi (CVVor) and EICV yield the most complete performance and best accuracy across all the investigated test problems (see Fig.~\ref{fig:NMRSE}). They achieve the smallest NRMSE values, close to zero for all test functions, with the updated surrogate model (i.e., updated with 25 samples). An important difference between these adaptive techniques is the computational effort. The computational effort required by methods based on the LOOCV error such as CVVor or EICV is relatively higher, especially as the dimensions increase. However, they yield accurate results and require less development and user-knowledge. Depending on the complexity and dimension of the problem, the SFEICV technique may be useful in the first few adaptively selected samples, but as more samples are collected, the EICV method becomes more promising. This can be seen in Figs.~\ref{fig:NMRSE}(a)(d)(e)(f) where SFEICV yields similar NRMSE values to EICV after 10 adaptively selected training points, i.e., 30 total training points, whereas, EICV results in better improvement in the surrogate model accuracy after 30 samples. 

Note that two separate GP models are constructed in EICV and SFEICV (which are special cases of the general method proposed in Section \ref{sec:learningfunc}), thus they require more computational effort than the other adaptive sampling techniques that are not based on the LOOCV error. However, they do not require the tessellation of the input space into Voronoi cells, which also requires high effort, as in CVVor and LOLA. In addition, the accuracy of EICV and SFEICV is better or comparable to the other methods for different benchmark functions.

\section{Engineering Application}\label{Sec:Numerical example}

\begin{figure}[h]
\centering
    \includegraphics[width=.125\textwidth]{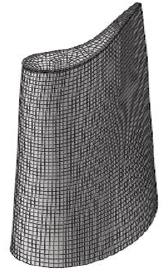}
  \caption{FEM model of engine rotor blade}
  \label{fig:fem_model}
\end{figure}

A simplified gas turbine engine blade model (see Fig. \ref{fig:fem_model}) is studied in this section to demonstrate the proposed surrogate modeling approach for an engineering application with high-dimensional spatio-temporal output. Six output quantities from the physics model (creep equivalent strain (CEEQ), creep damage (Dc), von Mises stress (Mises), and displacements in x, y, and z directions) are of interest, as shown in Table~\ref{table:inpout}; these are available at a large number of spatial and temporal points. The input space is four-dimensional and consists of turbine blade coating thickness (TBC thickness), turbine output rate (T1T rate), and transient and constant creep rates. The ranges of the inputs are: TBC thickness [2/3*nominal,4/3*nominal], T1T rate [80*nominal,100*nominal], transient and constant creep rate [-3$\sigma$,3$\sigma$]. These model inputs are considered as uncertain. The boundary conditions are assumed to be known with certainty. The total number of nodes is 29,374 for each output quantity of interest (QoI) and the total number of time steps is 54. Thus, the problem has over 176,244 spatial dimensions (i.e., 29,374 $\times$ 6) in the output space for a single FEM simulation. Considering also the temporal dimension, the problem dimension is over 9,000,000 (i.e., 176,244 $\times$ 54) just for a single FEM simulation.
\begin{table}[h]
    \caption{Physics model input and output parameters}
    \label{table:inpout}
\centering{%
    \resizebox{.48\textwidth}{!}{%
        \begin{tabular}{@{\extracolsep{\fill}}l*{2}l@{\extracolsep{\fill}}}
        \hline\noalign{\smallskip}
            Input & Values  \\
            \noalign{\smallskip}\hline\noalign{\smallskip}
            TBC thickness & $\frac{2}{3} \times$ nominal to $\frac{4}{3} \times$ nominal \\
            T1T rate & 80 $\times$ nominal to 100 $\times$ nominal \\
            Transient creep rate & -3$\sigma$ to 3$\sigma$ \\
            Constant creep rate & -3$\sigma$ to 3$\sigma$ \\
            \hline\noalign{\smallskip}
            Output & Nomenclature  \\
            \noalign{\smallskip}\hline\noalign{\smallskip}
            Creep equivalent strain & CEEQ \\
            Creep damage & Dc \\
            Von Mises stress & Mises \\
            X displacement & X \\
            Y displacement & Y \\
            Z displacement & Z \\
        \noalign{\smallskip}\hline
        \end{tabular}
    }%
}%
\end{table}
\subsection{Generation of Initial Training Data}
The initial surrogate model is built with 46 training points (from 46 FEM runs). Prior to this study, nine initial input settings had been generated using Taguchi L9 orthogonal design \cite{Taguchi1989} for exploratory analysis. To this set of 9 points, we added 37 points, generated using maximin Latin hypercube sampling (LHS). These 46 sets of inputs and the corresponding FEM outputs are then used to build the initial surrogate model.

In order to obtain the maximin LHS for the 4-dimensional input space, 100 LHS designs are generated. Based on the spread of the training points in the design space, the design with the highest spatial coverage is selected as the optimum. To quantify how spread out the points are in each design set, the sum of pair-wise Minkowski distances when p = 1 within the set is selected as a metric. For two row vectors, $\mathbf{x} \in \mathbb{R}^n$ and $\mathbf{y} \in \mathbb{R}^n$ in the matrix, the Minkowski distance is defined as:
\begin{equation}
    d(\mathbf{x},\mathbf{y})=\big(\sum_{i=1}^n \lvert\mathbf{x}_i- \mathbf{y}_i\rvert^p\big)^{(1/p)}.
\end{equation}
The higher the $d(\mathbf{x},\mathbf{y})$ is, the higher the occupancy rate (coverage) in the design space. The sum of the pair-wise Minkowski distances was calculated for each LHS design and the optimum design is selected to give the 37 additional training points for the initial surrogate model.

\subsection{Dimension Reduction of the Output Space}
Based on the correlation analysis, CEEQ and Dc are found to be perfectly correlated with each other at all time steps \cite{kapusuzoglu2022dimension}. Similarly, Y and Z displacements are negatively correlated, and the correlation becomes more significant at higher time steps. The features obtained in single-physics problem like structural dynamics have physical meaning, but in multi-physics problems where different output quantities are combined together, the physical meaning is not clear. In this work, dimension reduction is performed using rSVD on the entire multi-response output data in order to take this correlation between the output QoIs into account and improve the computational efficiency. Thus the obtained features represent all the QoIs. Then a \emph{single} surrogate model, Extra-Trees regressor, is built to predict multivariate time series for all 6 QoIs. 

To be able to jointly predict the QoIs at each time step for untested configurations, the output data matrix $\mathbf{S}$ is constructed as follows:
\begin{small}
\begin{equation}
    \mathbf{S}=[\mathbf{D}_1 \mathbf{D}_2 \mathbf{D}_3 \mathbf{D}_4 \mathbf{D}_5 \mathbf{D}_6]_{(n_{time}\times n_{sim}) \times (n_{nodes}\times n_{QoI})}
\end{equation}
\end{small}
\noindent where $\mathbf{D}_i\in$ $\mathbb{R}^{n_{time}\times n_{nodes}}$, $\ i=(1,2,...,n_{QoI})$ is the $i$th QoI data matrix.

The matrix $\mathbf{S} \in$ $\mathbb{R}^{(n_{sim}\times n_{time})\times(n_{nodes}\times n_{QoI})}$ is constructed such that the number of rows is equal to the total number simulations ($n_{sim}$) times the total number of time steps of each simulation ($n_{time}$), and the number of columns is equal to number of nodes ($n_{nodes}$) times the total number of QoIs ($n_{QoI}$). For example, for the final surrogate model we have output data for 66 simulations and the output data matrix $\mathbf{S}_{(54\times 66)\times (29374\times 6)}$ consists of approximately 629 million data points. 

We have used RMSE to evaluate the reconstruction accuracy and the percentage of variance explained by the top important features for further analysis. The percentage of variance explained by the top 50 and 20 features in the first and second feature spaces, respectively, is 99\% and the RMSE value for reconstruction error is approximately 0.0006, which is small compared to the average magnitude of the predicted quantities (0.01). We have not used the MAE metric for calculating the reconstruction error. MAE will allow the engineers to choose a different threshold that has a physical meaning for each individual quantity of interest. However, the values of MAE thresholds for reconstruction error in individual quantities are not known to the engineer. Whereas, when we use RMSE, we can quantify one error measure for all the quantities together and assess the quality of reconstruction.

\subsection{Surrogate Model Construction and Cross-Validation}
After the important features are identified, several types of surrogate models are investigated as discussed in Section \ref{sec:surrmodel} and based on the cross-validation results given in Table.~\ref{table:cv_results}, the Extra-Trees regressor is observed to give the best results with an average RMSE value of 0.10513 (15\% better than the next best model). We have used 7-fold in Table 2 to partition the data evenly (k=7 is a divisor of the sample size). We tried both with 5-fold and 10-fold. The cross-validation results were almost identical. The LOOCV error is used to evaluate the surrogate model accuracy. The inputs to the surrogate model are TBC thickness, T1T rate, transient and constant creep rates and the outputs of the surrogate model are the top 20 important features. The proposed approach is able to distinguish between errors stemming from bad approximation of the surrogate model and errors stemming from insufficient quality of the project space. We have not distinguished them in our numerical example, but have included both of them together in our analysis, especially since the reconstruction error was negligible compared to the surrogate model error.

\begin{table}[h]
    \caption{7-fold cross-validation results in terms of RMSE}
    \label{table:cv_results}
    
\centering{%
    \resizebox{.485\textwidth}{!}{%
        \begin{tabular}{@{\extracolsep{\fill}}l*{7}c@{\extracolsep{\fill}}}
        \hline\noalign{\smallskip}
            & \multicolumn{6}{c}{Model} \\
            \cline{2-7}
            Fold \# & Extra-Trees & LightGBM & XGBoost & CatBoost & DNN & GP \\
            \noalign{\smallskip}\hline\noalign{\smallskip}
            1 & 0.1722 & 0.2029 & 0.1823 & 0.1780 & 0.2102 & 0.2205 \\
            2 & 0.1885 & 0.2146 & 0.2113 & 0.1843 & 0.2243 & 0.2315 \\
            3 & 0.1329 & 0.1829 & 0.1732 & 0.1362 & 0.1856 & 0.1795 \\
            4 & 0.1090 & 0.1257 & 0.1219 & 0.1155 & 0.1351 & 0.1583 \\
            5 & 0.1123 & 0.1286 & 0.1199 & 0.1136 & 0.1209 & 0.1348 \\
            6 & 0.1039 & 0.1273 & 0.1166 & 0.1065 & 0.1385 & 0.1421 \\
            7 & 0.1426 & 0.1835 & 0.1462 & 0.1503 & 0.1857 & 0.1812 \\
            Average & 0.1373 & 0.1665 & 0.1531 & 0.1406 & 0.1714 & 0.1782 \\
        \noalign{\smallskip}\hline
        \end{tabular}
    }%
}%
\end{table}

\begin{figure}[h]
\centering
    \includegraphics[width=.25\textwidth]{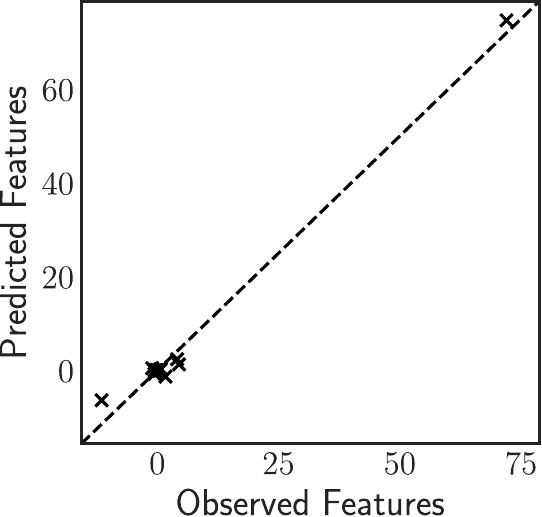}
  \caption{Predicted vs. observed values of important features for FEM run 23}
  \label{fig:cv-ext}
\end{figure}

\begin{figure}[h]
\centering
    \includegraphics[width=.5\textwidth]{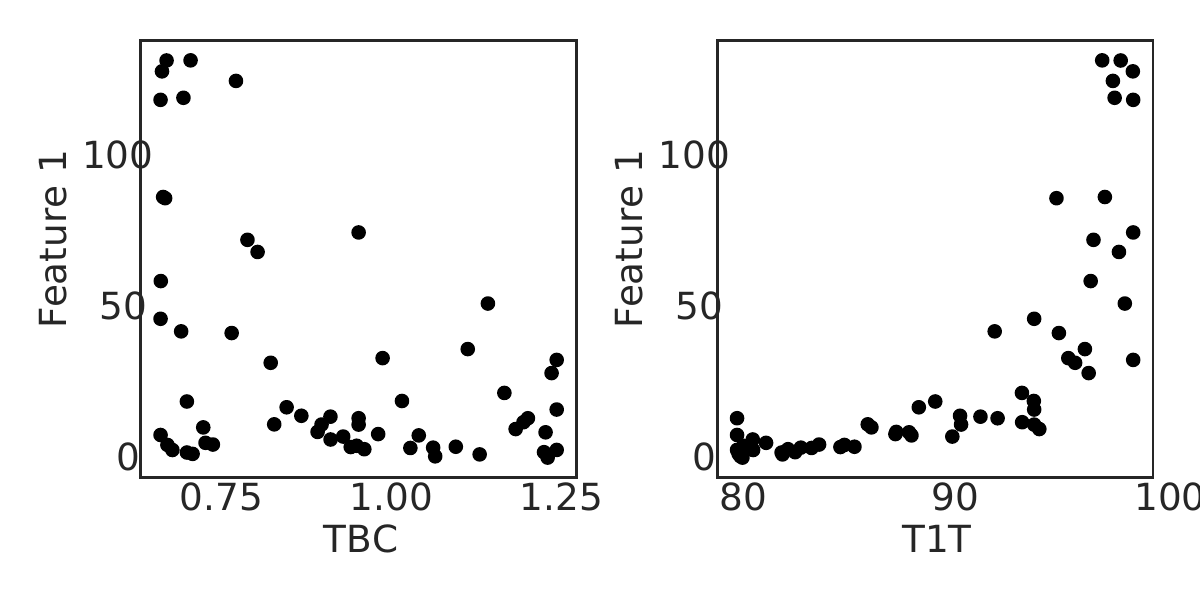}
  \caption{Actual Feature 1 values vs. TBC thickness and T1T rate}
  \label{fig:inp_feat1}
\end{figure}

For illustration purposes, the predicted feature values versus the actual feature values are shown in Fig.~\ref{fig:cv-ext} for FEM run 23. Note that the pairs of observations and predictions are close to the 45-degree line, showing good agreement. The $R^2$ and RMSE values of the predictions shown in Fig.~\ref{fig:cv-ext} are 0.97 (quantifies how close the observed vs. predicted data are to the 45-degree line) and 0.45 (small error compared to the actual feature values), respectively. Figure~\ref{fig:inp_feat1} shows actual Feature 1 vs. the input parameters TBC thickness and T1T rate. Although the general trend shows a decrease of Feature 1 values with an increase in TBC thickness, there are still small Feature 1 values across all values of TBC thickness. Feature 1 values increase as T1T rate increases, indicating a possible quadratic relationship. From the physics of the problem, it is known that T1T values greater than 90 result in strong creep cases (i.e., large CEEQ and Dc values) and Feature 1 values corresponding to these cases are significantly larger than Feature 1 values corresponding to the weak creep cases (i.e., T1T $<$ 90). Moreover, the correlations between the model inputs and the top 5 important features are computed (not shown in the paper due to space limitation), and the most significant correlation is found to be between T1T rate and Feature 1, with a value of 0.85.

\subsection{Adaptive Surrogate Modeling}\label{sec:adap_app}
\begin{figure}[h]
\centering
    \includegraphics[width=.5\textwidth]{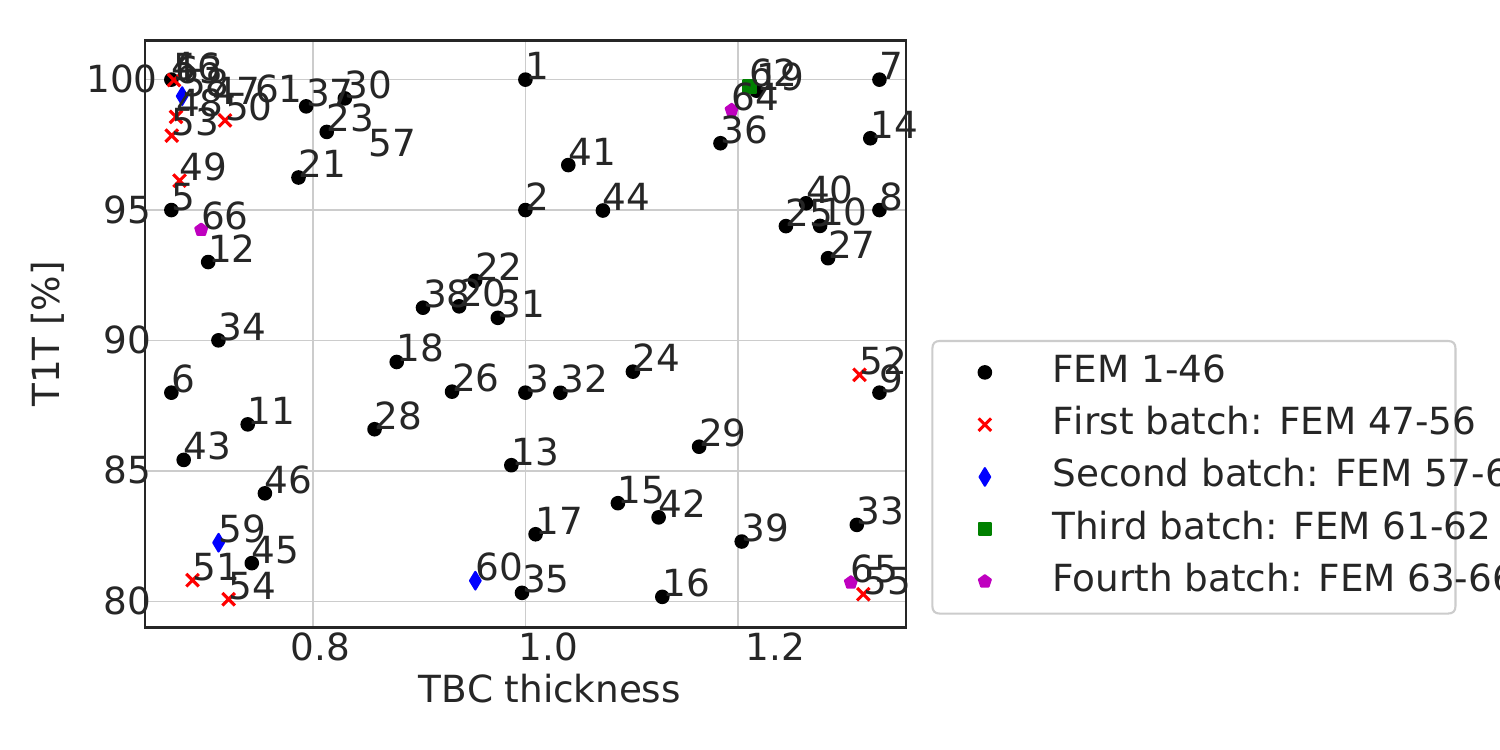}
  \caption{Initial FEM training data and adaptively selected additional samples}
  \label{fig:allfemcases}
\end{figure}

Next, the training points corresponding to additional FEM runs (47-66) are adaptively selected based on the proposed methodology of adaptive training point selection. These runs are completed in 4 batches (see Fig.~\ref{fig:allfemcases} and Table~\ref{table:batches}). From the surrogate model improvement point of view, it is more desirable to select one training point at a time. However, this is not possible in industrial settings due to the fact that running a single FEM model at each step is inefficient use of human and computational resources.
\begin{table}[h]
    \caption{Optimal hyperparameter values of each batch}
    \label{table:batches}
\centering{%
    \resizebox{.48\textwidth}{!}{%
        \begin{tabular}{@{\extracolsep{\fill}}l*{8}c@{\extracolsep{\fill}}}
        \hline\noalign{\smallskip}
            Batch id & $\beta_1$ & $\beta_2$ & $\alpha$ & $\gamma$ \\
        \noalign{\smallskip}\hline\noalign{\smallskip}
            First batch: FEM 47-56 & 0.30 & 0.03 & 1 & 1 \\
            Second batch: FEM 57-60 & 0.56 & 0.23 & 1 & 1 \\
            Third batch: FEM 61-62 & 1 & 1 & 1 & 1 \\
            Fourth batch: FEM 63-66 & 0.53 & 0.04 & 1 & 1 \\
        \noalign{\smallskip}\hline
        \end{tabular}
    }%
}%
\end{table}

A more efficient strategy in this application was to run between 2 and 10 FEM models at the same time. Thus, multiple new training points were proposed in each batch where the number of points for each batch is chosen based on error quantification, space filling, as well as expert opinion regarding inputs of interest. The simplest strategy would be to rank the candidate points and select the desired number of best points. However, this is not ideal since it does not consider the information overlapping of the new points in that batch, which can lead to clustered batch points \cite{liu2018survey}. (This is because the space-filling component of the learning function in Section \ref{sec:learningfunc} only gives the score for each single candidate point in terms of its distance from the existing points; whereas when we consider multiple points in a batch, we also have to consider the distances among them). In our case, the selection of the new training points in each batch considers how \emph{informative} and \emph{diverse} they are \cite{liu2018survey}, from the perspectives of both geometry and physical behavior. This way not only are the new points sampled in the regions of interest but they are also far away from each other.

Efficient global optimization (EGO) techniques that select the next points in batches extended the expected improvement (EI) such that EI is maximized when multiple points are added to the data set \cite{henkenjohann2007efficient,ponweiser2008clustered}. However, the optimization of the multiple point learning functions is computationally challenging \cite{ginsbourger2010kriging}. Here we considered a simpler and computationally cheaper approach (i.e., multiple points were selected based on their learning function values out of all candidate points) for selecting the next points in batches. For example, in the second batch, only four (57-60) samples were selected that satisfied the criteria of informativeness and diversity based on the proposed learning function values out of all candidate points. Similarly, in the third and fourth batches, only two points each were selected for the same reasons.

The first batch of new training points are selected by performing the proposed methodology described in Algorithm~\ref{algorithm} on the previous 46 FEM runs, and the second batch of new training points are obtained by doing the same analysis on the previous 56 FEM runs (i.e., 46 initial points plus 10 points from the first batch of additional training points), and so on. For most of the adaptively sampled points, i.e., FEM 47-60, and 63-66, the parameters of the learning function are estimated by using the two-step algorithm described in Section \ref{sec:learningfunc}. The expert opinion is only used for some of the training points to guide the learning. For example, when the relative error was more important for the engineers, we have adjusted beta in an ad-hoc fashion. Most of these samples are from regions where the absolute error has high values (i.e., physical quantities take medium to high values, which is of interest to the analyst in this case). However, for the problem of interest it is also equally important to be able to accurately predict physical quantities that are close to zero. (For example, in the gas turbine engine, some input values might lead to low creep; such cases could be quickly screened and ignored in further detailed analysis). Thus, for a few of the adaptively selected additional samples such as FEM 61 and 62 the parameters of the learning function shown in Eq.~\eqref{eq:learningfunc} are chosen as unity to give equal weight to all metrics in order to facilitate choosing more diverse training points, including from regions where the relative error is higher, within each batch (e.g., FEM 64 and 65).

\begin{figure*}[h]
\centering
    \includegraphics[width=.7\textwidth]{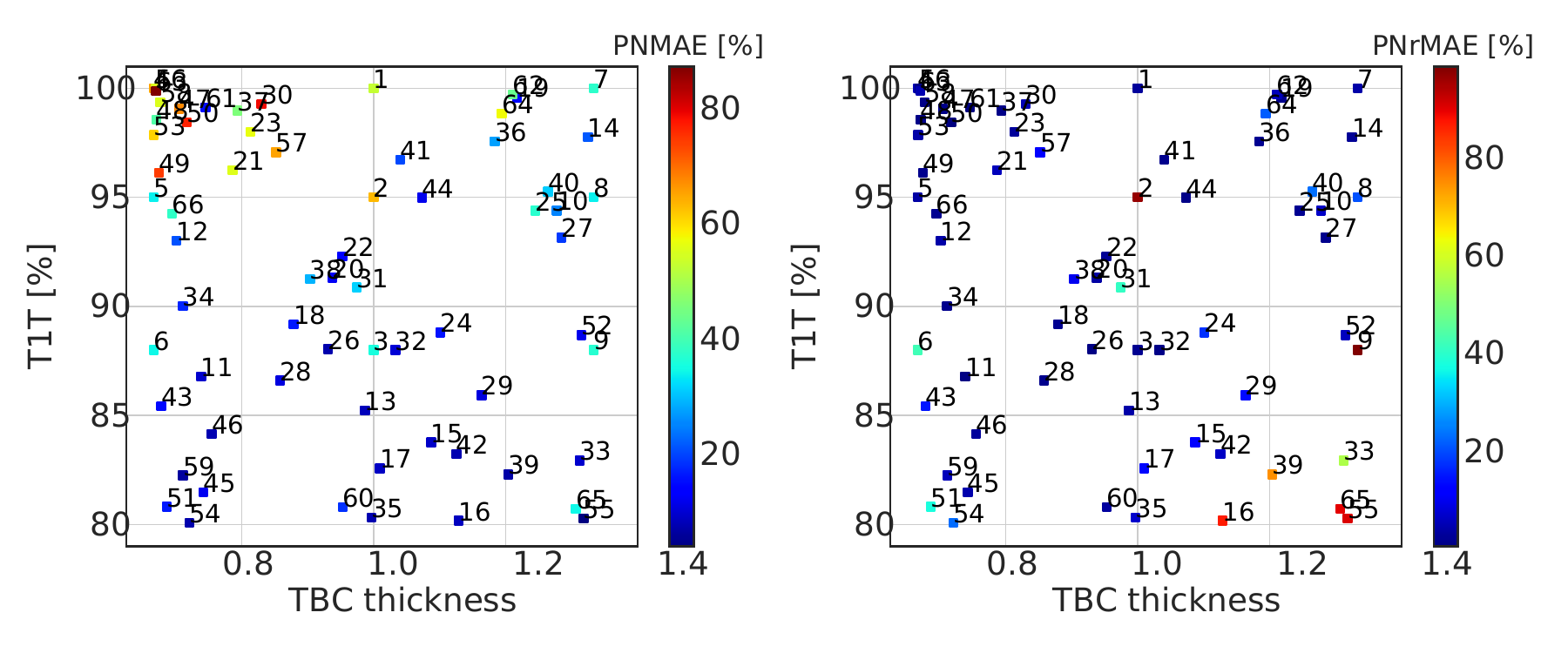}
  \caption{Final surrogate model prediction accuracy in the original space for leave-one-out FEM runs}
  \label{fig:pnmae_pnrmae}
\end{figure*}

The surrogate model prediction accuracy is further analyzed in the original space by mapping the surrogate model predictions in the features space back to the original space. As discussed earlier, the percentage of nodes having MAE $>$ \emph{threshold} (PNMAE) and the percentage of nodes having rMAE $>$ \emph{unity} (PNrMAE) are defined for error analysis. The threshold value of 0.00025 is chosen based on expert opinion. The accuracy of the final surrogate model is shown in Fig.~\ref{fig:pnmae_pnrmae} using the two error metrics PNMAE and PNrMAE in a two-dimensional space, i.e., TBC thickness vs T1T rate. The error metrics have different values in different regions. For example, the absolute error metric, PNMAE, has smaller values in the bottom half region, where the physical quantities are close to zero, and has larger values in the top left region, where the physical quantities take medium to high values. Whereas, the relative error of the predicted creep response, PNrMAE, has greater values in regions where weak creep behavior is dominant. This is caused by dividing with a very small number. Since the physical quantities are close to zero in weak creep regions (i.e., the bottom half region T1T $< 90$) the denominator of Eq.~\eqref{eq:rmae} would be close to zero. 

\begin{figure*}[h]
\centering
    \includegraphics[width=.8\textwidth]{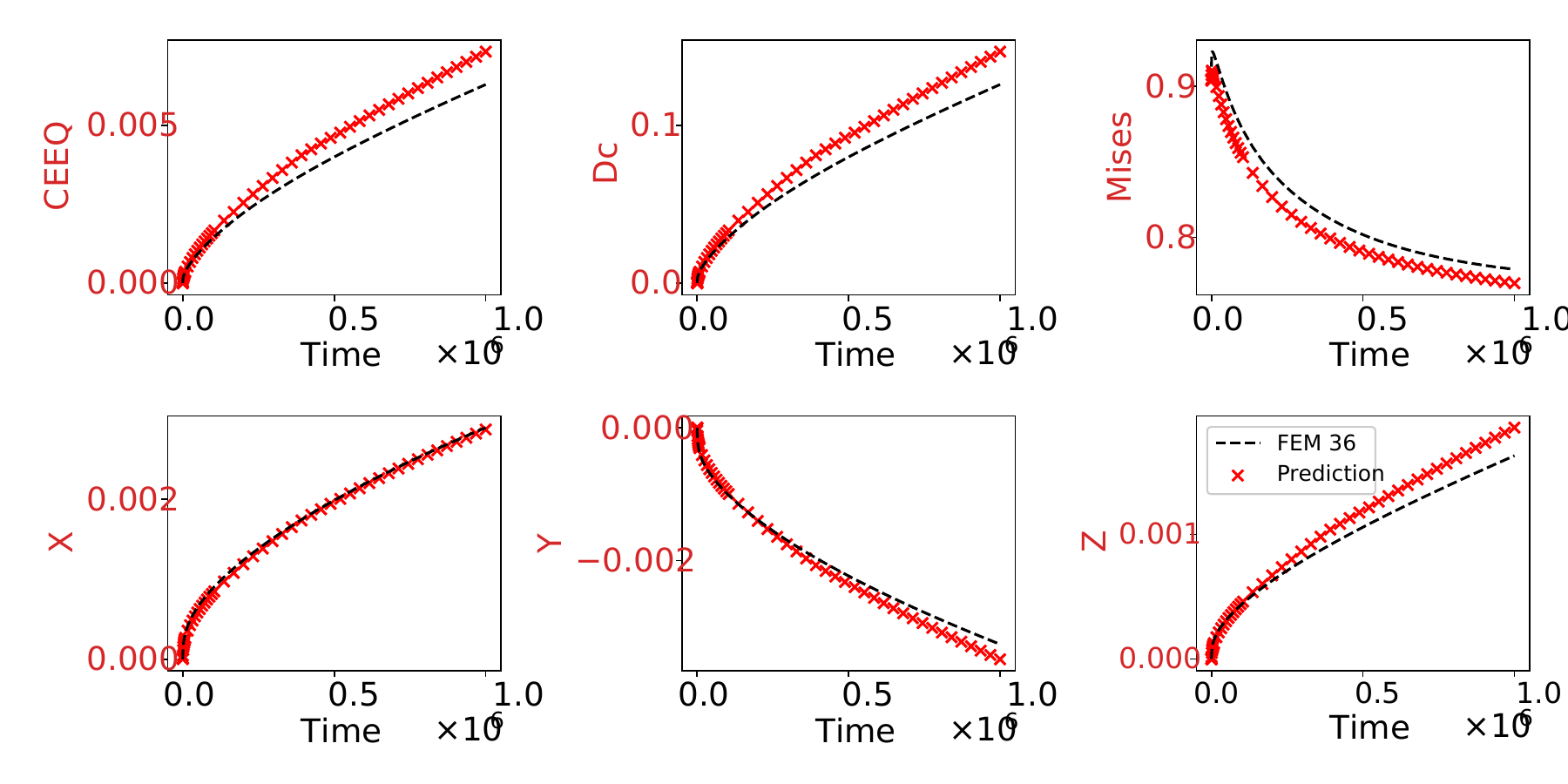}
  \caption{Final surrogate model prediction compared against FEM run 36 at node 5057}
  \label{fig:pred_vs_fem37}
\end{figure*}

The final surrogate model predictions after 66 training points at an important node (id 5057: hot spot, based on analyst's knowledge) are compared against the FEM results of run 36, where high creep damage is observed, as shown in Fig.~\ref{fig:pred_vs_fem37}. It is observed that the predictions are in good agreement with the FEM results at all time steps (similar trends in some cases if not actual values), which reflects the effectiveness of the proposed methodology.

The LOOCV results are shown in Table~\ref{table:LOOCVmetrics} using three different error metrics. In order to have a fair comparison, the improvement in the surrogate model accuracy is assessed on the same data set (i.e., first 46 FEM runs). The columns of Table~\ref{table:LOOCVmetrics} represent the average LOOCV error values on the first 46 FEM runs. In general, the accuracy of the surrogate model is observed to improve with each batch of adaptively selected samples. The average RMSE value, which is obtained by averaging the RMSE values obtained for each LOOCV fold, decreases with each new batch of adaptively selected training points. For example, it decreases from 0.00254 to 0.00222 with the addition of 10 new adaptively selected training points (samples 47-56) based on the LOOCV on the first 46 FEM runs. Similarly, PNMAE and PNrMAE values decrease as we include adaptively selected training points during the surrogate model construction process. 
\begin{table}[h]
    \caption{Average LOOCV results of the first 46 FEM runs in terms of PNMAE, PNrMAE, and RMSE}
    \label{table:LOOCVmetrics}
\centering{%
    \resizebox{.48\textwidth}{!}{%
        \begin{tabular}{@{\extracolsep{\fill}}l*{7}c@{\extracolsep{\fill}}}
        \hline\noalign{\smallskip}
            Training data & Avg. PNMAE & Avg. PNrMAE & Avg. RMSE \\
        \noalign{\smallskip}\hline\noalign{\smallskip}
            FEM 1-46 & 27.9 & 19.3 & 0.00254 \\
            FEM 1-56 & 26.5 & 17.6 & 0.00222 \\
            FEM 1-60 & 26.3 & 17.3 & 0.00219 \\
            FEM 1-62 & 24.9 & 16.1 & 0.00209 \\
            FEM 1-66 & 24.6 & 15.8 & 0.00193 \\
        \noalign{\smallskip}\hline
        \end{tabular}
    }%
}%
\end{table}

The number of training points $N$ and the surrogate model accuracy are used as stopping criteria. The sequential adaptive sampling procedure is stopped after 66 training points, $N=66$. The adaptively improved surrogate model does not successfully reach the desired level of accuracy of average PNrMAE $\le$ 15\% (chosen based on expert opinion). However, the average PNrMAE value is 15.8\% after 66 training points (for the first initial 46 FEM runs), which suggests that with a few more FEM runs the desired level of accuracy could be reached. In addition to the average PNrMAE value being distinctly close to the desired level of accuracy, the PNMAE and RMSE values are also acceptable for the analyst after 66 training points. The largest contributor to the overhead cost is the adaptive sampling procedure. The second largest contributor is the validation of the surrogate model for each QoI using the appropriate error metric. All calculations except the FEM simulations are performed on a single desktop computer with an Intel 8-core CPU, 3.00 GHz base frequency, and 8 GB memory.

\section{Conclusion}\label{Sec:Conclusion}

This paper developed an approach for adaptive surrogate model construction in engineering problems with high-dimensional spatio-temporal output. The important features are obtained by performing a two-step dimension reduction using rSVD to map the original high-dimensional spatio-temporal output to an uncorrelated space. Cross-validation error is used to identify the most accurate surrogate model type. Once the most accurate surrogate model type is identified for the problem of interest, the prediction error in the original space is evaluated using LOOCV with different error metrics (i.e., PNMAE and PNrMAE). The subsequent adaptive sampling technique combines exploration and exploitation for adaptive improvement of surrogate model accuracy with the fewest possible runs of the expensive physics-based original model.

The proposed adaptive sampling technique is first compared with some of the existing adaptive schemes (Section~\ref{sec:proposedmethodcomparison}) using several test problems. The proposed method yields very good results across all investigated test problems. The effectiveness of the proposed methodology is next demonstrated for a turbine blade example (Section \ref{Sec:Numerical example}), using a time-dependent multi-physics dynamic system model with high-dimensional spatio-temporal outputs. It is observed that the adaptively improved surrogate model reaches an acceptable level of accuracy using the proposed strategy.

Future work can explore other large-scale engineering applications in order to further evaluate the performance of the proposed method, gain more insight about the method’s behavior for different problems, and further improve the methodology. The proposed adaptive sampling technique can be extended in future to include parallelization of sequential sampling for global optimization. Future work can also isolate the reconstruction error from the overall prediction error in the original space, which includes both the reconstruction error and surrogate model error.

\section*{Acknowledgments}

This study was funded by Mitsubishi Heavy Industries in Nagasaki and Takasago, Japan. The support is gratefully acknowledged. In addition, we acknowledge our paper in the AIAA SCITECH 2022 Forum \cite{kapusuzoglu2022dimension}; note that in that paper, we only presented preliminary results but did not discuss the methodology.

\section*{Declarations}

The authors declare that they have no known competing financial interests or personal relationships that could have appeared to influence the work reported in this paper.

\section*{Conflict of Interest}
The authors declare that they have no conflict of interest. 

\section*{Replication of Results}
Unfortunately, we cannot share the in-house code used in this work because of confidential proprietary nature as identified by the sponsor.








\begin{appendices}

\section{Benchmark Test Functions}\label{app:1}

Two-dimensional benchmark functions used in this paper to evaluate the performance of the proposed method and compare its performance to existing adaptive surrogate model improvement methods are defined in Table~\ref{tab:testproblems}.

\begin{table*}[h]
\caption{Test problems}
\begin{center}
\begin{minipage}{340pt}
\begin{adjustbox}{width=.95\textwidth}
\begin{tabular}{@{}ll@{}}
    \toprule
    $\begin{aligned}
        &\text{Branin function} \\
        &y = \left[x_2 - 5\left(\frac{x_1}{2\pi}\right)^2 + \frac{5x_1}{\pi} -6 \right]+10\left(1-\frac{1}{8\pi} \right)\text{cos}(x_1)+10 \\
        &x_1 \in [-5,10],\ x_2 \in [0,15]
    \end{aligned}$ & \raisebox{-.3\totalheight}{\includegraphics[width=0.3\textwidth]{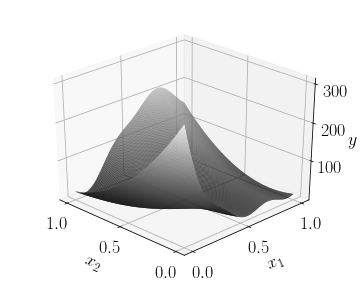}}\\
    $\begin{aligned}
        &\text{Goldstein \& Price function} \\
        &y = \left[1+\left(x_1+x_2+1\right)^2\left(19-14x_1+3x^2_1-14x_2+6x_1x_2+3x^2_2\right)\right] \\ \nonumber 
        &\quad \times \left[30+\left(2x_1-3x_2\right)^2\left(18-32x_1+12x^2_1+48x_2-36x_1x_2+27x^2_2\right)\right]\\ \nonumber
        &x_1,\ x_2 \in [-2,2]
    \end{aligned}$ & \raisebox{-.3\totalheight}{\includegraphics[width=.3\textwidth]{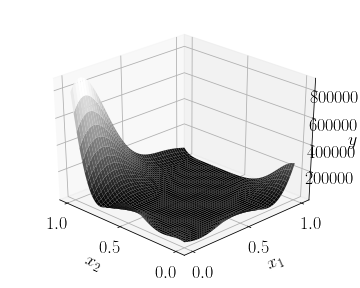}}\\
    $\begin{aligned}
        &\text{Sasena's function} \\
        &y = 2+0.01(x_2-x_1^2)^2 + (1-x_1)^2+\\
        &\quad 2(2-x_2)+7\text{sin}(0.5x_1)\ \text{sin}(0.7x_1x_2)\\ \nonumber
        &x_1,\ x_2 \in [0,5]
    \end{aligned}$ & \raisebox{-.3\totalheight}{\includegraphics[width=.3\textwidth]{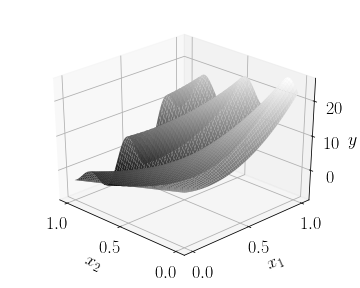}}\\
    $\begin{aligned}
        &\text{Alpine function} \\
        &y = \text{sin}x_1\ \text{sin}x_2 \sqrt{(x_1x_2)}\\ \nonumber
    &x_1,\ x_2 \in [0,10]
    \end{aligned}$ & \raisebox{-.3\totalheight}{\includegraphics[width=.3\textwidth]{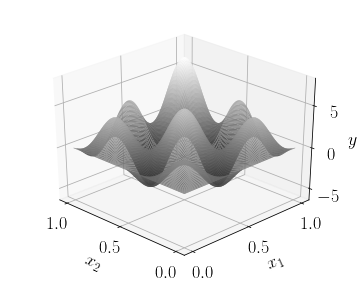}}\\
    $\begin{aligned}
        &\text{Modified form of function in Meckesheimer et al.~\cite{meckesheimer2001computationally}} \\
        &y = e^{(x_1-x_2)^2}+e^{(10-x_1)^2}-x_1x_2\\ \nonumber
    &x_1,\ x_2 \in [0,10]
    \end{aligned}$ & \raisebox{-.3\totalheight}{\includegraphics[width=.3\textwidth]{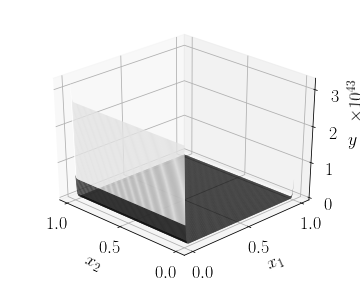}}\\
    $\begin{aligned}
        &\text{Modified form of function in Jin et al.~\cite{jin2002sequential}} \\
        &y = \text{cos}(10x_1^2) +3.1\lvert x_1-0.7\rvert+2x_1^2+\text{sin}\left(\frac{1}{\lvert x_1-0.7\rvert+0.31}\right) +2x_2^2 \\ \nonumber
    &x_1,\ x_2 \in [0,1]
    \end{aligned}$ & \raisebox{-.3\totalheight}{\includegraphics[width=.3\textwidth]{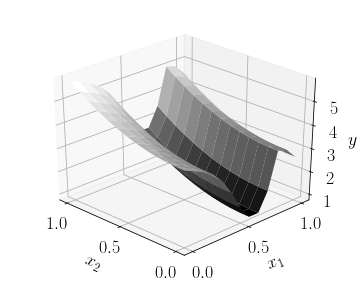}}\\
    $\begin{aligned}
        &\text{Schwefel function} \\
        &y = 418.9829n - \sum(x_i\ \text{sin}\sqrt{\lvert x_i\rvert}\\ \nonumber
        &x_i \in [-500,500],\ i = 1, ..., n, n=2
    \end{aligned}$ & \raisebox{-.3\totalheight}{\includegraphics[width=.3\textwidth]{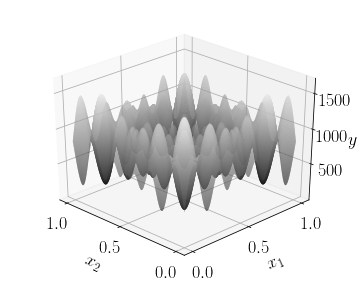}}\\
\end{tabular}
\end{adjustbox}
\end{minipage}
\end{center}
\label{tab:testproblems}
\end{table*}

\end{appendices}

\bibliographystyle{unsrt}  
\bibliography{references}

\end{document}